\documentclass{article}
\usepackage{graphicx}
\graphicspath{%
    {converted_graphics/}
    {/}
}
\begin{document}

\begin{center}
{\LARGE An Experimental Apparatus for Observing Deterministic}\vskip6pt

{\LARGE \ Structure Formation in Plate-on-Pedestal Ice Crystal Growth}\vskip%
12pt

{\Large Kenneth G. Libbrecht}\vskip4pt

{\large Department of Physics, California Institute of Technology}\vskip-1pt

{\large Pasadena, California 91125}\vskip-1pt

\vskip18pt

\hrule\vskip1pt \hrule\vskip14pt
\end{center}

\textbf{Abstract.} We describe an experimental apparatus for making detailed
morphological observations of the growth of isolated plate-like ice crystals
from water vapor. Each crystal develops a plate-on-pedestal (POP) geometry,
in which a large, thin, plate-like crystal grows out from the top edge of an
initially prismatic seed crystal resting on a substrate. With the POP
geometry, the substrate is not in contact with the growing plate (except at
its center), so substrate interactions do not adversely affect the crystal
growth. By controlling the temperature and supersaturation around the
crystal, we can manipulate the resulting ice growth behavior in predictable
ways, producing morphologies spanning the full range from simple faceted
hexagonal plates to complex dendritic structures. We believe that the
experimental apparatus described here will allow unprecedented
investigations of ice crystal growth behaviors under controlled conditions,
identifying and exploring robust morphological features in detail. Such
investigations will provide valuable observational inputs for developing
numerical modeling techniques that can accurately reproduce the faceted and
branched structures that frequently emerge during diffusion-limited crystal
growth.

\section{Introduction}

The growth of crystalline ice Ih from water vapor near the triple point has
long been a fascinating case study for understanding the physical processes
governing crystal growth (for a review, see \cite{libbrechtreview05}). The
emerging growth morphologies are largely deterministic, yet highly variable
with changing temperature and supersaturation around the crystal. In the
pure system of ice and water vapor (with no background gas), measured growth
rates as a function of temperature, supersaturation, and facet surface are
controlled mainly by attachment kinetics at the ice surface. Quantitative
measurements thus provide insights into the structure of the ice surface,
including surface premelting and step energies on facet surfaces, which are
both poorly understood at present \cite{kglalphas13}. In the presence of an
inert background gas, particle transport also plays an important role, and
the ice system can be used for quantitative investigations of the
diffusion-limited growth of structures that are both branched and faceted 
\cite{reiter05, gg09, kglca13, kelly13}. Recent measurements suggest that
structure-dependent attachment kinetics (SDAK) is also a key factor
determining the observed ice growth behavior \cite{sdak03, sdak12}.

Ukichiro Nakaya pioneered the study of ice crystal growth from water vapor
in the 1930s, producing isolated crystals on rabbit hair \cite{nakaya54}.
Numerous researchers have subsequently grown ice crystals on various thin
fibers, on capillary tubes, on various substrates, and even on the ends of
electrically enhanced ice needles (see \cite{libbrechtreview05} and
references therein). Each experimental method has its own merits for making
quantitative growth measurements as well as qualitative morphology
observations that reveal different aspects of ice growth behavior.

We recently described the formation of \textquotedblleft
plate-on-pedestal\textquotedblright\ (POP) ice crystal growth, in which a
thin plate-like crystal grows out from a small prismatic seed crystal
resting on a substrate. Notably, the growing ice plate forms entirely above
the substrate, being held up only by a small ice pedestal. As described in 
\cite{sdak12}, this unusual structure is especially useful for exploring the
SDAK instability, although growth on ice needles shows perhaps even greater
promise in this regard \cite{kglca13, kgldual14}.

To our knowledge, the POP method for growing ice crystals was first
described by Gonda, Nakahara, and Sei in the 1990s \cite{gonda90, gonda97},
although the authors did not explicitly point out the POP structure of their
ice crystals. It appears that little subsequent work has been done to
further develop the POP method. The present paper picks up where these
earlier works left off, seeking to develop some of the unrealized potential
of the plate-on-pedestal method.

Beyond its application in scientific investigations of crystal growth
dynamics, the POP\ method also presents considerable opportunity to further
the art of ice crystal growth from water vapor. In that direction we have
been exploring various techniques to grow ice crystals that resemble
atmospheric snow crystals, exhibiting a broad variety of branched and
faceted structures with their distinctive six-fold symmetry. Time-lapse
photography of the POP\ crystals also reveals many aspects of snow-crystal
growth behavior with unprecedented clarity.

\begin{figure}[tbp] 
  \centering
  \includegraphics[width=5in,keepaspectratio]{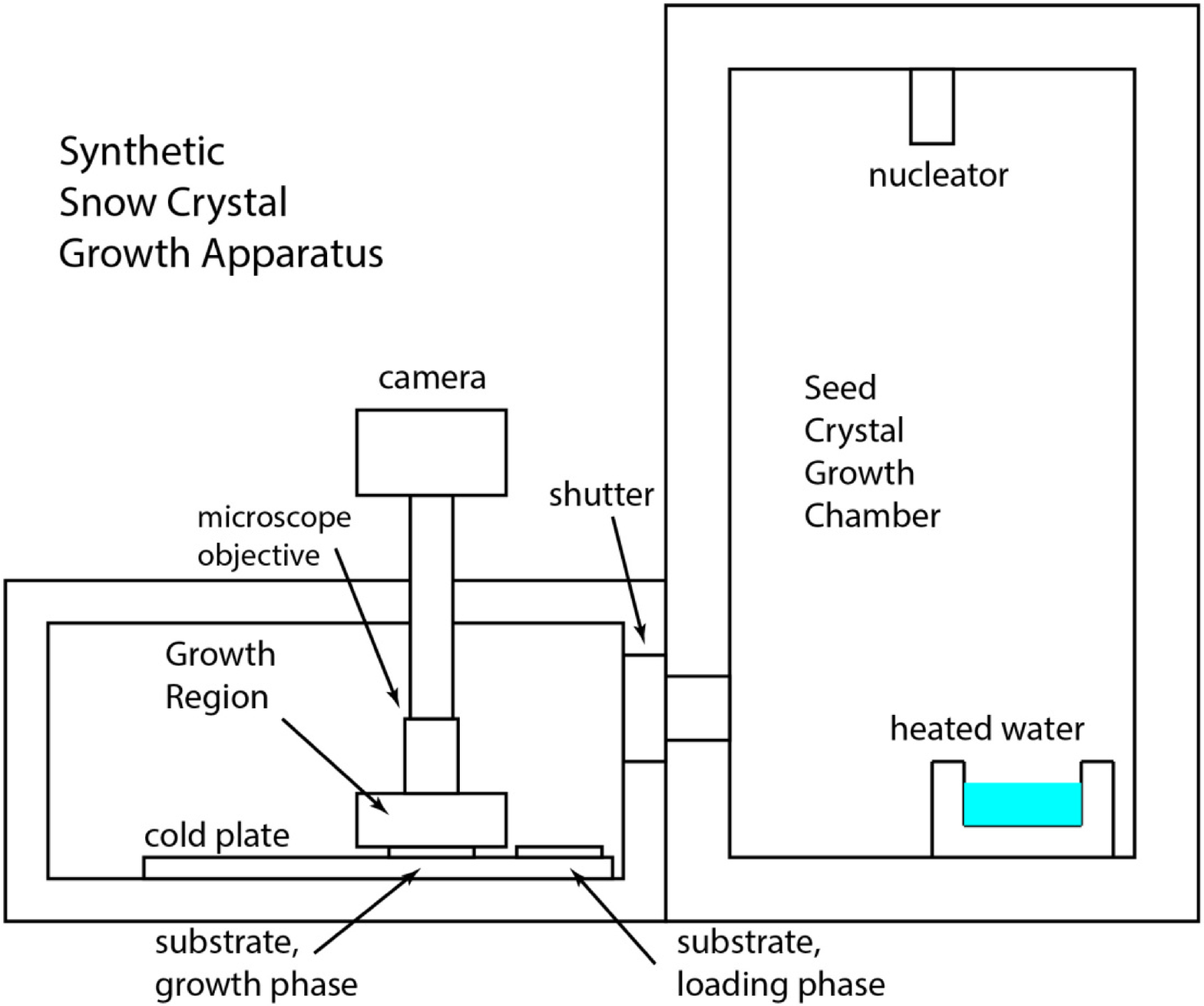}
  \caption{A schematic diagram of our
apparatus for growing and photographing synthetic snow crystals, described
in detail in the text.}
  \label{largeview}
\end{figure}

\section{Experimental Apparatus}

Figure \ref{largeview} shows an overview of the hardware we use for growing
ice crystals from water vapor in a background gas of air. During a typical
run, freely floating seed crystals are generated continuously in the seed
crystal growth chamber (see Figure \ref{largeview}), as described below. To
produce a growing POP\ crystal, a shutter is opened to allow some of the
seed crystals to fall randomly onto the substrate in its loading position.
The substrate is then moved to the growth region, where it is positioned to
place an isolated seed crystal at the center of the microscope field of
view. Moist air is then blown gently onto the substrate, causing the test
crystal to grow. The temperature and supersaturation around the crystal can
be changed as it grows, while a camera records its development. We now
examine the various parts of this apparatus in detail.

\subsection{The Seed Crystal Growth Chamber}

Seed crystals are produced in a custom-built cold chamber with inside
dimensions of approximately 40x40x100 cm, the large dimension being the
height. The chamber is cooled by a laboratory chiller (SP Scientific/FTS
RS33-LT) that circulates temperature-regulated coolant (methanol) through
the walls of the chamber as well as the adjoining cold plate (see Figure \ref%
{largeview}). Ordinary laboratory air fills the seed chamber, which does not
have an air-tight seal. The chamber is insulated with approximately 5 cm of
styrofoam sheeting covering all the outer surfaces.

An insulated container containing one liter of ordinary tap water rests on
the bottom of the seed chamber, as shown in Figure \ref{largeview}, and the
water temperature is kept constant by a temperature regulator using an
immersed water heating element and water temperature sensor. The top of the
container is open to the air, so water evaporates and is carried by
convection to the rest of the chamber. As we have described previously \cite%
{libbrechtmorrisonfaber08, libbrechtarnold09}, convective mixing of the air
produces a fairly uniform temperature and supersaturation within the seed
chamber, except immediately above the water reservoir and near the chamber
walls (which soon become covered with frost, lowering the nearby
supersaturation). Water vapor is continually removed from the air by growing
ice crystals and by frost depositing on the walls of the chamber, and this
water vapor is continually replenished by evaporation from the water
reservoir. The air temperature is easily measured by placing a thermistor
inside the chamber. The supersaturation is more difficult to determine, but
can be inferred by the morphology of the growing crystals. We therefore
adjust the water temperature and chiller temperature until the desired
chamber air temperature and seed crystal morphology are produced.

We typically set the chiller temperature to -19 C and the water temperature
to 17 C, as this yields a continuous supply of thin, hexagonal plate-like
seed crystals with diameters in the 20-50 micron range. The air temperature
inside the seed crystal growth chamber remains near -15 C. Higher water
temperatures (while maintaining the air temperature near -15 C) yield higher
supersaturations and more dendritic structures. Thicker plates can be
produced with different air temperatures, while temperatures near -5 C yield
columnar crystals. See \cite{libbrechtmorrisonfaber08, libbrechtarnold09}
for additional details on free-fall growth chambers and the crystal
morphologies that result.

The nucleator at the top of the chamber consists of a small stainless steel
chamber (a standard 1.33-inch conflat nipple) with an interior volume of 
\symbol{126}25 cm$^{3}$ that is connected to a solenoid valve. The nucleator
assembly is inside the growth chamber, so its temperature is approximately
-15 C during operation. Pressurized room air flows into the nucleator
through a constriction that limits the flow. The flow rate is slow enough
that the air temperature becomes roughly equilibrated inside the nucleator.
Likewise, frost condenses on the inner walls of the nucleator, so the water
vapor content of the air is near the saturated value. Every ten seconds the
solenoid valve is opened, causing the pressurized air to rapidly expand and
enter the growth chamber. The rapid expansion produces a small amount of air
that is sufficiently cooled to nucleate ice crystals. Air pressures as low
as 15 psi will nucleate crystals, and we usually operate with 30 psi. Water
buildup inside the nucleator is removed after each run by operating it for
several hours when the chamber is at room temperature. With no initial ice
buildup, the nucleator can run continuously for at least ten hours without
difficulty.

The nucleated crystals float freely as they grow, until they eventually
settle to the bottom of the chamber. The fall times are typically a few
minutes, depending on temperature and supersaturation. Pulsing the nucleator
valve open every ten seconds thus produces a steady-state in which roughly a
million seed crystals are growing inside the chamber at any given time (this
number being determined by a visual estimate of the typical spacing between
crystals floating inside the chamber during operation). Shining a bright
flashlight into the chamber reveals sparkles caused by reflections off the
crystal facets, and this is a convenient way to verify that seed crystals
are present. 

Compressed air for both the nucleator and the crystal growth region is
supplied by an ordinary workshop air compressor with a built-in storage tank
and regulator, which automatically maintains the required 30 psi air
pressure. The compressed air is passed through an oil filter, then an
activated charcoal filter (containing coconut charcoal) to remove remaining
chemical contaminants from the air, and then a fine-pore fiber filter to
remove any remaining charcoal dust.

An aperture in the side of the seed crystal growth chamber connects it to
the adjoining main growth chamber (the left side of Figure \ref{largeview}).
The cold plate within the main chamber is cooled using the same circulating
coolant that flows through the walls of the seed chamber. To grow an POP
crystal, the substrate is first moved into its loading position (see Figure %
\ref{largeview}) and a shutter is opened between the two chambers. Random
air flow between the chambers carries a small number of crystals into the
second chamber, and some of these fall onto the substrate, a process that
takes a few seconds. The substrate is then moved to a covered region within
the main growth chamber, and a suitably isolated seed crystal is centered
under the microscope for subsequent growth and observation.

\begin{figure}[t] 
  \centering
  \includegraphics[width=4.0in,keepaspectratio]{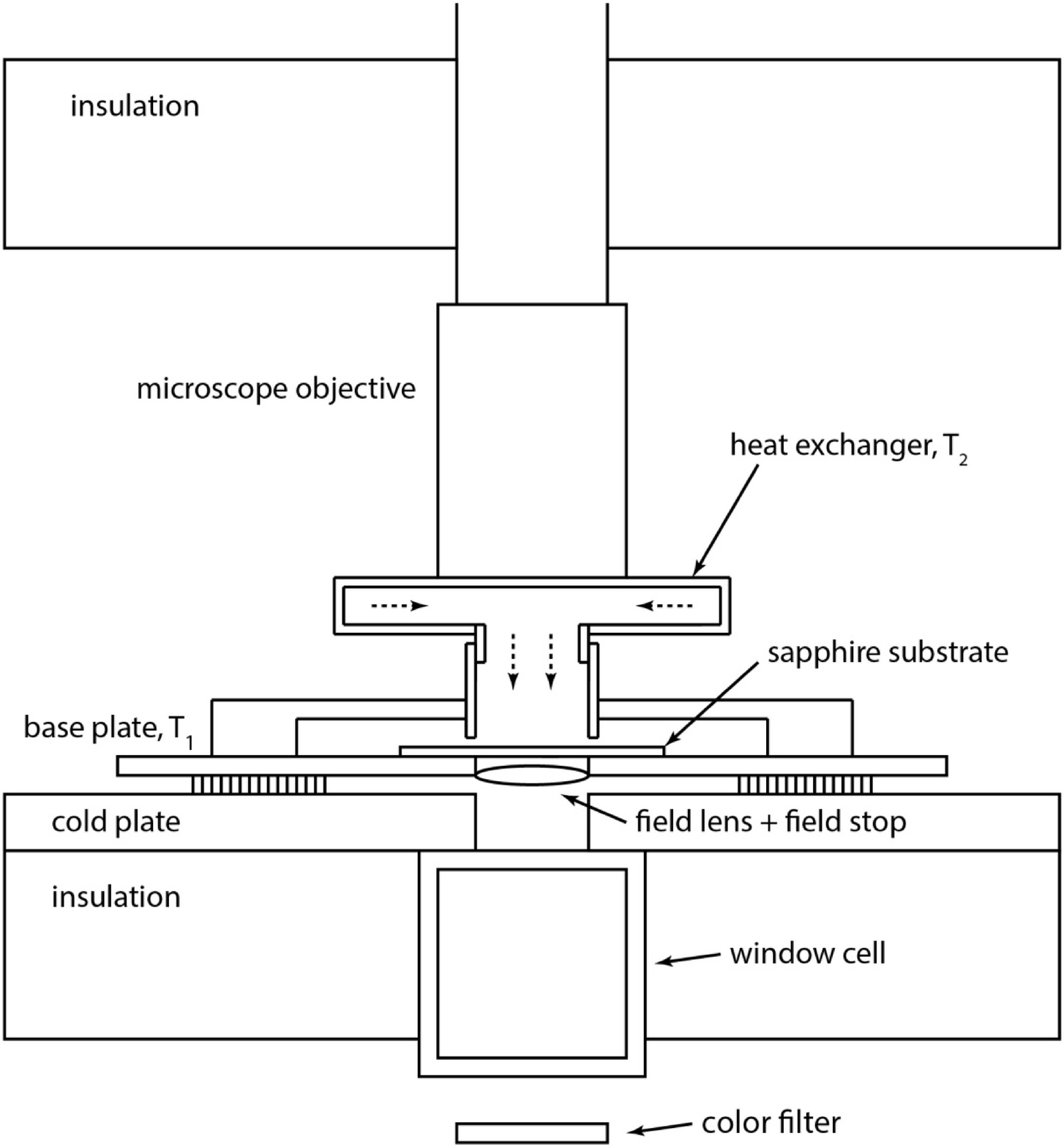}
  \caption{A schematic diagram of the
growth region shown in Figure \protect\ref{largeview}. Once a test crystal
has been positioned at the center of the microscope field-of-view, moist air
passing through the heat exchanger is blown onto the substrate, causing the
crystal to grow. The base-plate temperature $T_{1},$ the heat-exchanger
temperature $T_{2},$ and the air flow rate through the heat exchanger can
all be controlled as the crystal growth is observed.}
  \label{growthchamber}
\end{figure}

\subsection{The Growth Chamber}

Figure \ref{growthchamber} shows a more detailed view of the growth region
shown in Figure \ref{largeview}. The substrate is an uncoated sapphire disk,
50.8 mm in diameter and 1 mm thick, with the sapphire c-axis perpedicular to
the disk surface. The principal advantages to using sapphire in this
application are its high thermal conductivity and resistance to scratching.
The substrate slides on an anodized aluminum plate with a temperature $T_{1}$
maintained by a temperature controller (Arroyo Instruments model 5310) using
a thermistor temperature sensor (Cole-Parmer Digi-Sense 08491-15) in the
aluminum plate and thermoelectric heating/cooling modules beneath it. The
thermistors have an absolute accuracy better than $\pm $0.1 C, and the
temperature regulation is stable to better than $\pm $0.01 C under normal
operation. However, the temperature of the substrate and especially the air
immediately above it are not precisely equal to the aluminum plate
temperature, and this adds uncertainty to the ice crystal growth temperature.

The heat exchanger above the substrate is an aluminum plate at a temperature 
$T_{2}$ maintained by a separate temperature controller. Filtered room air
from the air compressor first passes through a baffled precooler kept near $%
T_{precool}=0$ C $(T_{precool}$ $>T_{2})$ to partially cool the air, and to
remove a large fraction of the water vapor initially present in the air. The
air then passes through a series of serpentine channels in the heat
exchanger plate before blowing down onto the substrate and the growing ice
crystal. The air flow rate is typically 200-300 ccm, measured using a
tapered-tube flow meter and controlled with a simple needle valve. This flow
rate replaces air in the guide tube (between the heat exchanger and the
substrate, as shown in Figure \ref{growthchamber}) about once per second.
The interior diameter of the guide tube is 1.6 cm, and its overall length
(from the bottom of the heat exchanger to the substrate) is approximately
2.3 cm. Air flows into the guide tube via four channels in the
heat-exchanger plate, arranged symmetrically around the circumference of the
top of the guide tube. The equal flow rates through the four input channels,
along with the cylindrical geometry of the guide tube assembly, were
engineered to produce a nearly cylindrically symmetric downward flow pattern
within the guide tube, with the flow axis centered on the growing crystal.
The guide tube temperature is kept near the substrate temperature $T_{1},$
and the guide tube is thermally isolated from the heat exchanger by a
thin-walled plastic tube.

During the intial cooldown of the apparatus, air is passed through the heat
exchanger for 30 minutes to deposit ice on its inner surfaces. The
temperature is set to $T_{2}<-20$ C during this time to make sure ice (and
not supercooled water) is deposited inside the heat exchanger. Once the heat
exchanger has been preconditioned in this way, air passing through it will
exit at temperature $T_{2}$ and be saturated with water vapor relative to
ice at $T_{2}.$ As it approaches the substrate, the air cools to near $%
T_{1}<T_{2}$ and thus becomes supersaturated. The degree of supersaturation
is controlled by setting the temperature difference $\Delta T=T_{2}-T_{1}$
as well as the air flow rate. To lowest order, the supersaturtion is
proportional to $\Delta T^{2}$ \cite{libbrechtreview05}. If the
supersaturation is sufficiently high, water droplets will condense on the
substrate around the growing ice crystal. In general, a higher $\Delta T$
and a higher flow rate produce a higher supersaturation around the growing
ice crystal. 

Modeling the temperature and supersaturation at the growing ice surface is
problematic with this apparatus for a number of reasons. The Reynolds number
of the flow is approximately 10, so the flow is probably not perfectly
laminar, and the timescale for the air in the guide tube to become
equlibrated via diffusion with the guide-tube walls is comparable to the
time it takes air to flow through the tube. Moreover, a stagnation point in
the flow occurs where the flow axis of the system intercepts the substrate
surface, at the position of the growing crystal, further complicating the
flow and thermal analysis. More importantly, water droplets condensing on
the substrate near the crystal substantially alter the supersaturation
field, and the amount of water condensation changes dramatically with
changes in $\Delta T$ and flow rate. The thermal connection between the edge
of a growing POP crystal and the substrate below is also a bit difficult to
determine accurately. For all these reasons, we do not expect that the
apparatus described here will be well suited for performing precision
measurements of ice growth rates under known conditions. It is better suited
for more qualitative studies examining ice crystal morphologies and growth
behaviors.

The microscope objective shown in Figure \ref{growthchamber} is part of the
heat exchanger assembly, but it is kept a few degrees warmer than $T_{2}$
using a heater dissipating 1-2 Watts into the objective body. This elevated
temperature is necessary to keep fog from condensing on the glass face of
the objective, which would interfere with optical imaging. We typically use
a Mitutoyo 5X Plan Apo objective with an achromatic reimaging lens (focal
length 250 mm) immediately behind it. This objective has a working distance
of 34 mm, a numerical aperture of 0.14, resolution of 2 $\mu $m, and a
depth-of-field of 14 $\mu $m. Focusing is done by moving the camera body (on
a StacShot rail), and some amount of focus stacking is typically needed for
optimal imaging of large crystals, owing to the shallow depth-of-field. The
image projects to about 1 $\mu $m per pixel of the sensor on our Canon EOS
5D camera.

The field lens shown in Figure \ref{growthchamber} reimages a color filter
onto the pupil of the objective for achieving Rheinberg illumination. With
this technique, the background in the image plane remains uniform regardless
of the color filter used. Our filters are not monochromatic, but have a
variety of color patterns. With the Rheinberg optical set-up, the color
filter provides different colors of light illuminating the crystal at
different input angles, while all colors illuminate the image plane
uniformly. The transmitted light is refracted through the crystal, which
acts as a complex lens. The birefringence and color dispersion of the ice
are both negligible in this optical arrangement. Using Rheinberg
illumination adds a sense of depth to the photographs, which helps
accentuate the internal structure and surface patterning of the growing ice
crystals. The window cell provides thermal insulation between the room and
the cold plate. A variety of baffles and constrictions (not shown in Figure %
\ref{growthchamber}) prevent condensation on the different optical
components while maintaining the desired temperature profiles during
operation. 

It should be emphasized that producing an isolated seed crystal on the
substrate is perhaps the most difficult step in using this apparatus. In
contrast, the growth process itself is quite straightforward. Seed crystals
fall randomly during loading, and their surface density on the substrate is
adjusted by how long the shutter remains open with the substrate in the
loading position. Also, many seed crystals are malformed and are therefore
not suitable candidates for further growth. If the surface density of seed
crystals is too low, it may not be possible to locate a well formed
specimen. If the density is too high, then it may not be possible to obtain
a sufficiently isolated crystal. Often several tries are needed to find a
suitable seed crystal, taking anywhere from 1-30 minutes. Between tries, the
substrate is heated to evaporate away the existing crystals.

We often use a heating trick to load larger and better formed seed crystals.
After warming the substrate to clean it, we set the substrate temperature
controller to the desired loading temperature $T_{1},$ but then begin the
loading process while the substrate temperature is still higher than $T_{1}.$
With proper timing, the warmer substrate tends to evaporate away smaller
crystals, while larger ones survive just long enough for the substrate
temperature to reach $T_{1},$ and at that point the remaining crystals are
stable. The end result is that smaller crystals are removed, leaving only
larger seed crystals behind.

When growing large, plate-like ice crystals, the ideal seed crystal is a
well-formed hexagonal plate with its basal surfaces parallel to the
substrate, and with no other crystals closer than at least 5 mm on the
substrate. The subsequent growth phase typically lasts 30-90 minutes and can
be recorded via the imaging system. The temperatures $T_{1}$ and $T_{2}$ are
adjusted with time (it requires about a minute for each to stabilize), along
with the flow rate, to obtain the desired growth behaviors. At the end of
the growth phase, the substrate is heated to just below 0 C so that the ice
crystals evaporate away, at which point the process can be started once
again. After a typical day-long run growing crystals, the entire system is
warmed to room temperature and baked to remove water.

Because a growing crystal is surrounded by air that has passed through the
heat exchanger, we took special care to reduce chemical contaminants in that
air. The charcoal filter was put into the air stream to remove contaminants,
and the fiber filter downstream from the charcoal filter was tested for odor
emission. The heat exchanger is baked at 40 C overnight before each run to
remove residual contaminants. We have not made quantitative measurements of
the remaining contaminant level, but we are able to grow quite thin ice
plates near -15 C. This observation is itself a good indication that the air
flowing into the growth region is sufficiently clean, since contaminants are
known to inhibit thin plate growth at this temperature \cite{chemical11}. 

\section{Investigations of Growing Crystals}

To date we have confined our crystal-growth investigations to temperatures
between $-5$ C and $-20$ C, mostly growing large, thin, plate-like crystals
with an overall plate-on-pedestal (POP) structure. It should be possible to
grow other morphologies using this same apparatus, but our current system
only allows imaging from the top, which is best suited for observing
plate-like crystals.

\subsection{Hexagonal Plates}

\begin{figure}[t] 
  \centering
  \includegraphics[width=6.0in,keepaspectratio]{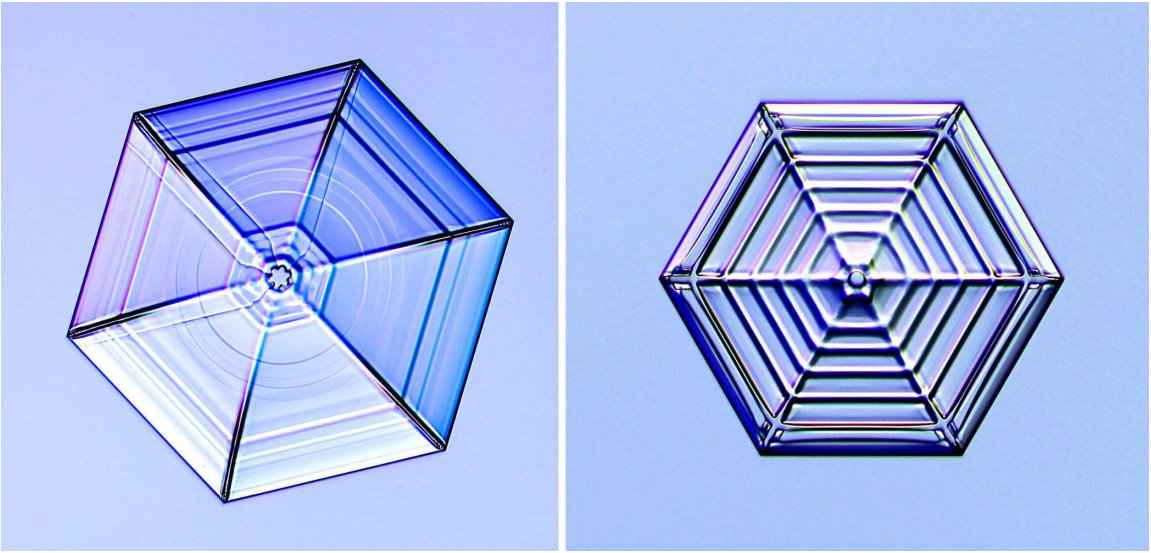}
  \caption{Photographs showing two
hexagonal POP ice crystals with several commonly seen morphological features
that are described in detail in the text. The crystal on the left exhibits
1) radial \textquotedblleft ridges\textquotedblright\ that divide the plate
into six sectors; 2) several concentric hexagonal \textquotedblleft
ribs\textquotedblright ; 3) several concentric circular macrosteps that grow
slowly inward; and 4) four curved dislocation lines (two of which flank the
upper left ridge). This crystal measures 1.5 mm from tip to tip. The crystal
on the right was subjected to periodic temperature changes that yielded a
spider's-web pattern of ridges and ribs. This crystal measures 0.75 mm from
tip to tip.}
  \label{simpleplate}
\end{figure}

Figure \ref{simpleplate} demonstrates several morphological features we
often see when growing hexagonal plate-like crystals. Focusing on the first
crystal in Figure \ref{simpleplate}, it was grown with $T_{1}=-15$ C and $%
T_{2}=-14$ C, so the ice temperature was near $-15$ C and the
supersaturation was below the water saturation level. The small star-shaped
figure at the center of the crystal outlines the pedestal that supports the
POP growth. The pedestal is approximately 70 $\mu $m in diameter, only
slightly larger than the diameter of the initial seed crystal. At
sufficiently low supersaturations, the seed crystal will grow out while the
edge remains in contact with the substrate. At temperatures within several
degrees of $-15$ C, however, even modest supersaturations yield the POP
morphology. All POP crystals thus have a small central pedestal, although it
is not always obvious in the photographs.

Note that six radial \textquotedblleft ridges\textquotedblright\ are clearly
visible in the first crystal in Figure \ref{simpleplate}, dividing the
hexagon into six sectors. Similar ridges are present in many POP crystals
growing near $-15$ C, and they are commonly seen in natural snow crystals as
well. The ridges originate at the corners of the hexagon and thus trace the
location of the corners as a function of time as the crystal grew. For
perfectly symmetrical growth, the ridges would all be straight. The growth
was not perfectly symmetrical with this crystal, so some of the ridges are
slightly curved. The origin of the asymmetry was some combination of
neighboring crystals (which affect the supersaturation in the air
surrounding the observed crystal), air flow asymmetries, and seed crystal
asymmetries (for example, tilt of the basal surfaces relative to the
substrate). The largest asymmetry is typically from neighboring crystals; in
the absence of neighbors, we can often grow crystals that are more
symmetrical than the first example in Figure \ref{simpleplate}.

The concentric hexagonal \textquotedblleft ribs\textquotedblright\ seen in
both crystals in Figure \ref{simpleplate} arise when the growth conditions
are altered with time. The ribs are essentially like tree rings that record
changes in the growth of the edge of the crystal as a function of time. With
constant external conditions, these ribs are largely absent. Whenever the
growth conditions are changing, however, ribs like those shown in these
examples are common morphological features.

The faint circular features seen in the first crystal Figure \ref%
{simpleplate} are inwardly growing macrosteps on the top surface of the
plate. These are best imaged using Rheinberg illumination, which highlights
such features. We have observed that the top surfaces of many POP crystals
are typically quite flat except for similar macrosteps. The ribs and ridges
are usually surface features confined to the lower surfaces of POP crystals
(the surface closer to the substrate). One goal of future 3-D computer
modeling efforts will be to reproduce these specific mophological features
in hexagonal POP crystals, as they are observed over a broad range of
conditions.

The first crystal in Figure \ref{simpleplate} also exhibits four curved
lines that arise from dislocations in the ice crystal. These lines are
unusual, suggesting that most simple plates are essentially dislocation
free. This particular crystal was allowed to branch (forming six separate
branches) early in its growth, and the branches later merged back together
to form a single plate. The dislocations appeared during these mergers, as
the branches did not recombine with atomic accuracy. Without the branching
step, plates typically do not exhibit dislocation lines.

\begin{figure}[t] 
  \centering
  \includegraphics[width=3.8in,keepaspectratio]{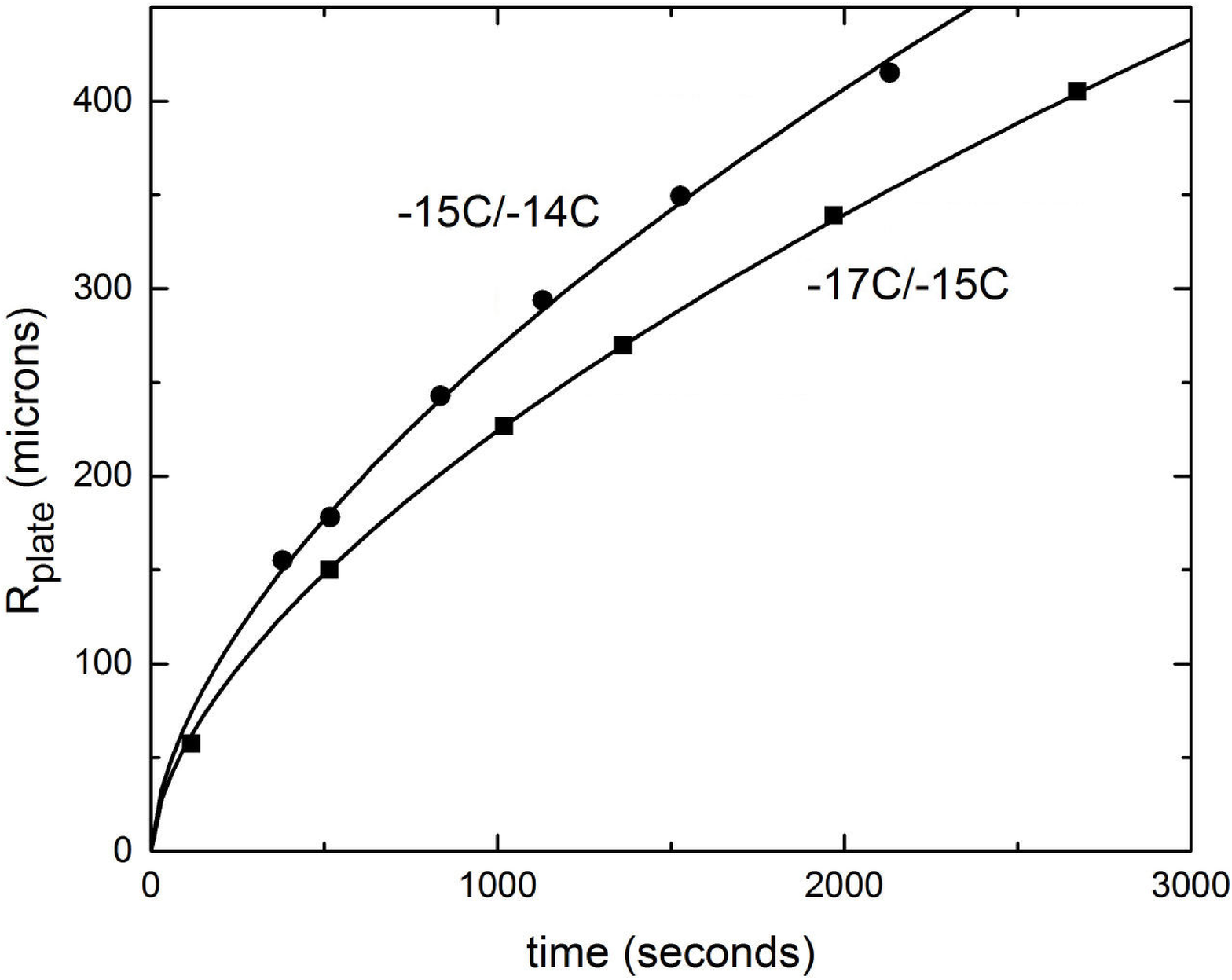}
  \caption{Example measurements of the
growth of two hexagonal-plate POP crystals. The plate radius is taken to be
half the tip-to-tip distance between opposite corners of each crystal. The
curves passing through the data points have the functional form $R=At^{0.6}.$
The labels indicate the temperature of the substrate $T_{1}$ (first number)
and the temperature of the heat exchanger $T_{2}$ (second number), which
remained constant as the crystals grew.}
  \label{growthrates}
\end{figure}

We have been able to grow hexagonal plates up to 2 mm in diameter rather
easily. The radius increases typically as $R\sim t^{\alpha },$ where $t$ is
the growth time and $\alpha $ is between $0.6$ and $0.7,$ depending on
initial conditions and other factors. Figure \ref{growthrates} shows some
typical growth rate data. Hexagonal plates grow quite readily at
temperatures between $-17$ C and $-13$ C.

\begin{figure}[tbp] 
  \centering
  \includegraphics[width=6.0in,keepaspectratio]{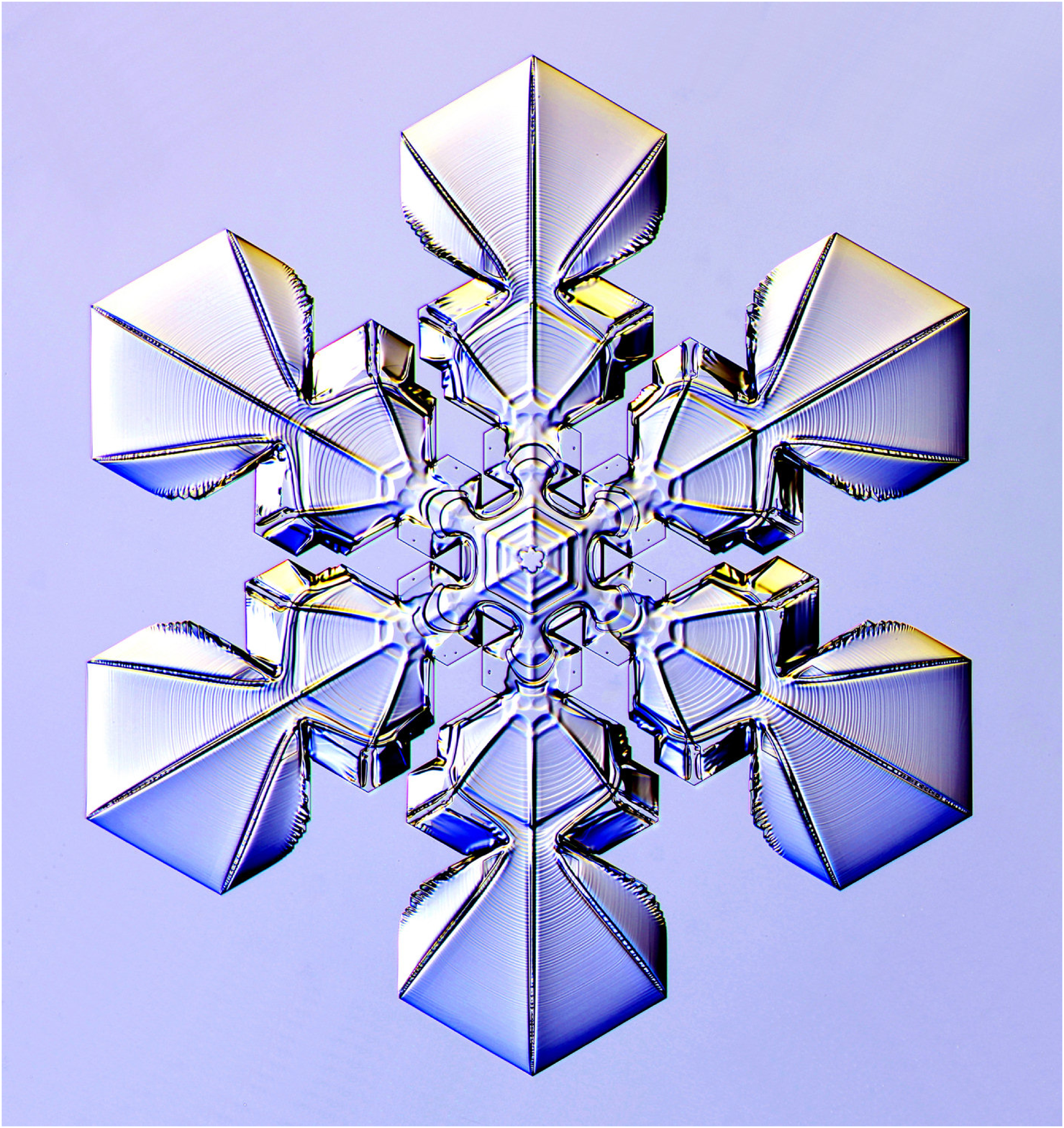}
  \caption{Photograph of a large POP
crystal that resembles sectored-plate snow crystals that grow in the
atmosphere. The size of this crystal is 3.3 mm from tip to tip. A variety of
ridges, ribs, and macrosteps are all visible in the photograph. The
irregularly shaped nub in the center of the crystal outlines the small
pedestal that supports the much larger plate above the substrate.}
  \label{sectoredplate}
\end{figure}

In \cite{sdak12} we presented a quantitative analysis of growing hexagonal
plates, using cellular automata modeling to extract information about the
attachment kinetics on the prism surface. Those measurements indicated the
importance of the SDAK instability in the growth of plate-like ice crystals.
The work described in the present paper supports that conclusion, but we
have not done additional quantitative investigations of the growing
crystals, focusing instead on morphological investigations and on techniques
for growing stellar snow crystals similar to those found in the atmosphere.

Going beyond simple hexagonal plates, Figure \ref{sectoredplate} shows a
more elaborate sectored-plate snow crystal made in this apparatus, which
exhibits many of the same morphological features seen in hexagonal plates.
Ridges, ribs, macrosteps, and faceted prism surfaces are all present in this
photograph, as they are to varying degrees in many of our POP crystals. By
examining these features in detail as a function of the applied growth
conditions, and by recording their formation and development as a function
of time, much can be done to further refine and develop computational
methods for simulating diffusion-limited crystal growth.

\subsection{Branching}

\begin{figure}[tbp] 
  \centering
  \includegraphics[width=5.9in,keepaspectratio]{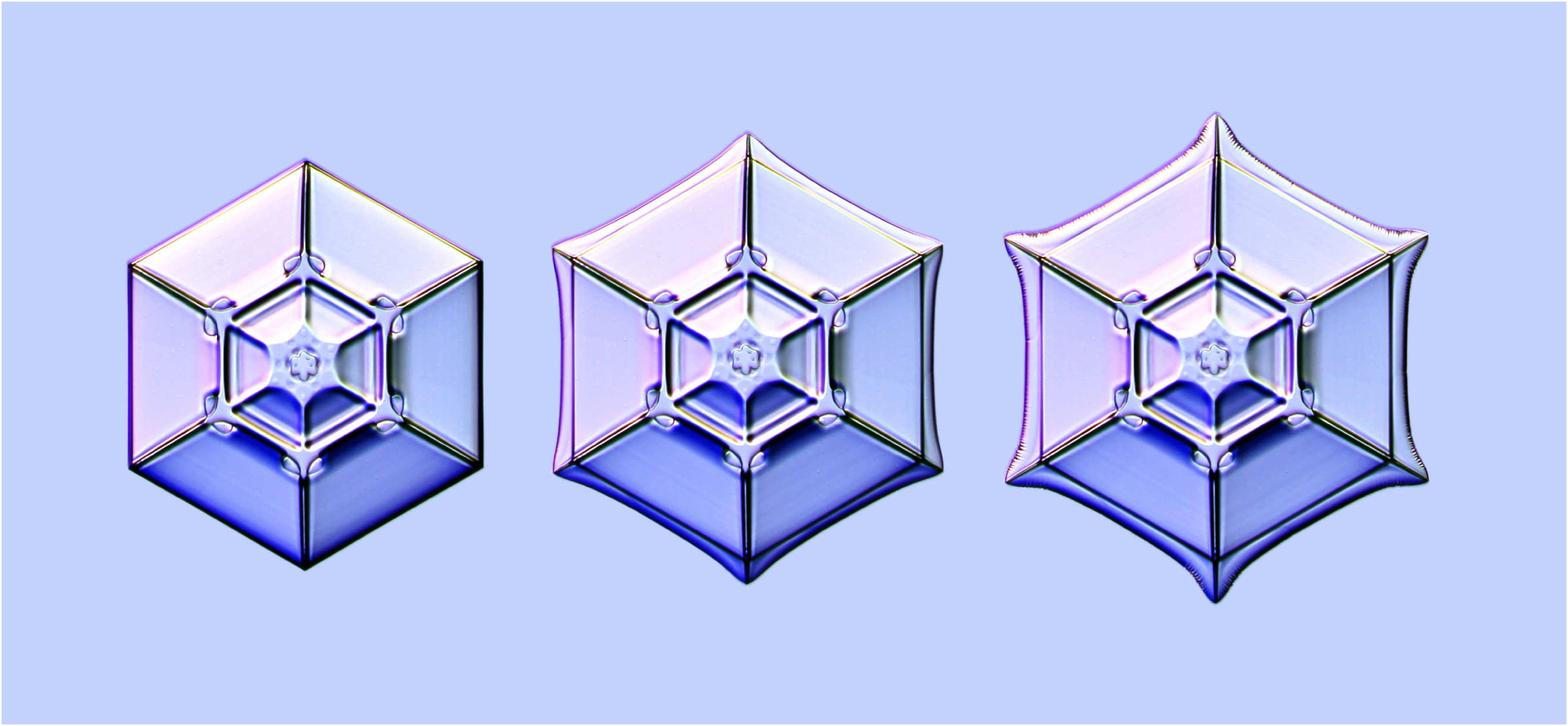}
  \caption{A series of photographs, taken
at different times, showing the formation of branches on a hexagonal POP
crystal. The crystal measured 0.6 mm from tip to tip in the first
photograph, when it was growing stably with $T_{1}=-13$ C and $T_{2}=-12$ C.
Reducing $T_{1}$ to $-15$ C increased the supersaturaion around the crystal,
initiating the development of branches as the corners of the hexagonal plate.}
  \label{branching}
\end{figure}

As is well known from numerous previous studies, branching via the
Mullins-Sekerka instability is important in snow crystal growth \cite%
{libbrechtreview05}, and an example of the branching instability in a POP
crystal is shown in Figure \ref{branching}. The first image in this figure
shows a hexagonal plate crystal while it was growing stably, maintaining its
overall hexagonal shape as it grew. Increasing $\Delta T,$ and thus the
supersaturation surrounding the crystal, initiated a branching event. Since
the tips of the hexagon projected farther out into the supersaturated air,
they accumulated more water vapor and grew more rapidly than other areas on
the crystal, causing the tips to project out even farther. The resulting
growth instability caused branches to sprout from the corners of the
crystal, as demonstrated in Figure \ref{branching}. Branching events such as
these are easily initiated and often observed in our POP crystals.

\begin{figure}[tbp] 
  \centering
  \includegraphics[width=6.0in,keepaspectratio]{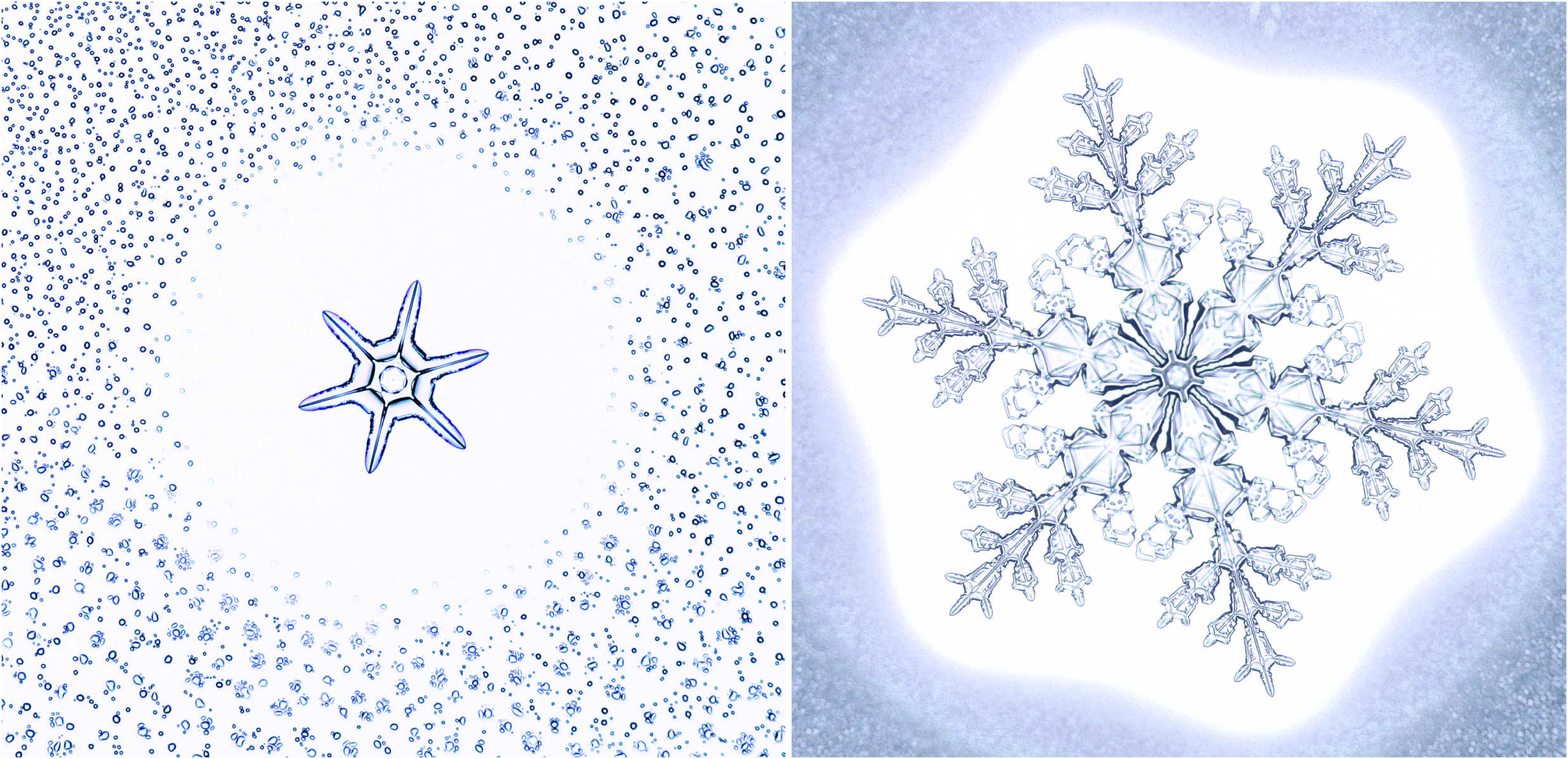}
  \caption{Two photographs of POP crystals
showing the deposition of water droplets on the substrate surrounding each
crystal. The crystals measured 0.36 mm (left) and 1.9 mm (right) from tip to
tip. Increasing either $\Delta T$ or the air flow rate generally results in
droplet depostion closer to the growing crystals.}
  \label{droplets}
\end{figure}

Increasing the supersaturation to initiate branching in POP crystals often
results in the condensation of water droplets on the substrate surface, as
shown in two examples in Figure \ref{droplets}. These droplets can remain
liquid for hours at temperatures near $-15$ C, and they prevent the
supersaturaion around the crystal (with respect to ice) from increasing
substantially above the water saturation point. Increasing the air flow rate
especially tends to cause droplet formation close to the growing crystal.
These droplets can be removed by reducing $\Delta T$ until they evaporate
away, but often they remain surrounding a growing crystal for long periods.
In some instances, they appear to stablilize the crystal growth and improve
the overall six-fold symmetry of the crystal. Often a neighboring crystal
(several millimeters away from the main POP crystal) will grow in such a way
that it intersects the substrate as it grows, eventually bringing it into
contact with some of the deposited droplets, causing them to freeze. The
frozen droplets soon contact neighboring droplets, and the front of frozen
droplets expands until the entire field is frozen, a process that typically
takes a few minutes. Once the central POP crystal is surrounded by a field
of frozen droplets, the supersaturation becomes quite low, inhibiting
subsequent growth.

\begin{figure}[t] 
  \centering
  \includegraphics[width=6.0in,keepaspectratio]{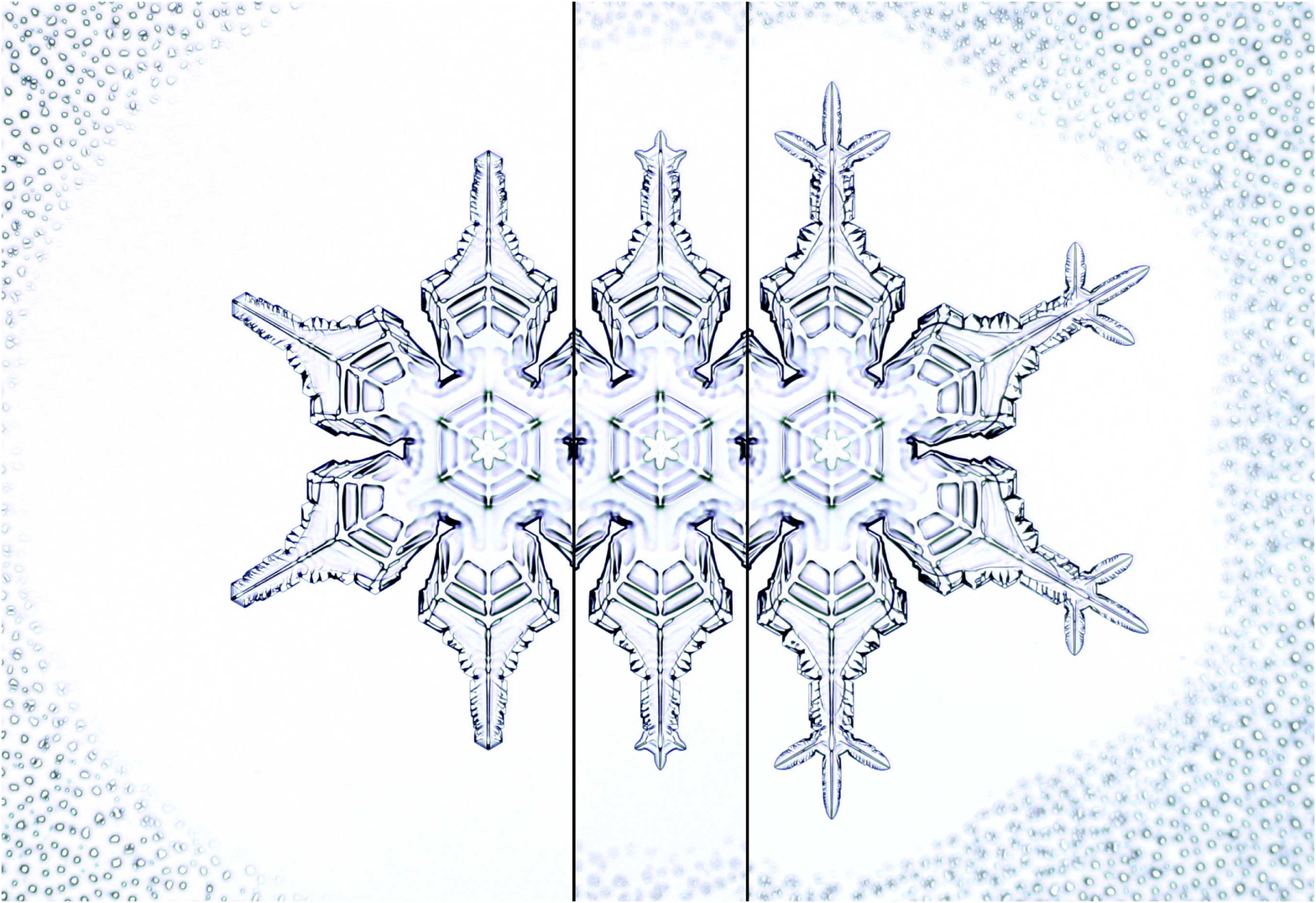}
  \caption{A composite image of a POP
crystal taken at three different times, showing an induced sidebranching
event. In the first image (left), the supersaturation was relatively low,
which caused the tips of the branches to slow their growth and become
faceted. In the second image, the supersaturation was increased, so water
droplets condensed more closely around the outer edges of the crystal. The
increased water supply caused sidebranches to sprout from the corners of the
arm tips. In the third image, the sidebranches have grown longer while the
supersaturation remained high. Note the symmetrical growth of the crystal,
and that the tips of the branches and sidebranches remain rounded when the
supersaturation is relatively high. Note also that most of the crystal
growth takes place at the outermost edges of the crystal.}
  \label{sidebranching}
\end{figure}

\begin{figure}[tbp] 
  \centering
  \includegraphics[width=6.0in,keepaspectratio]{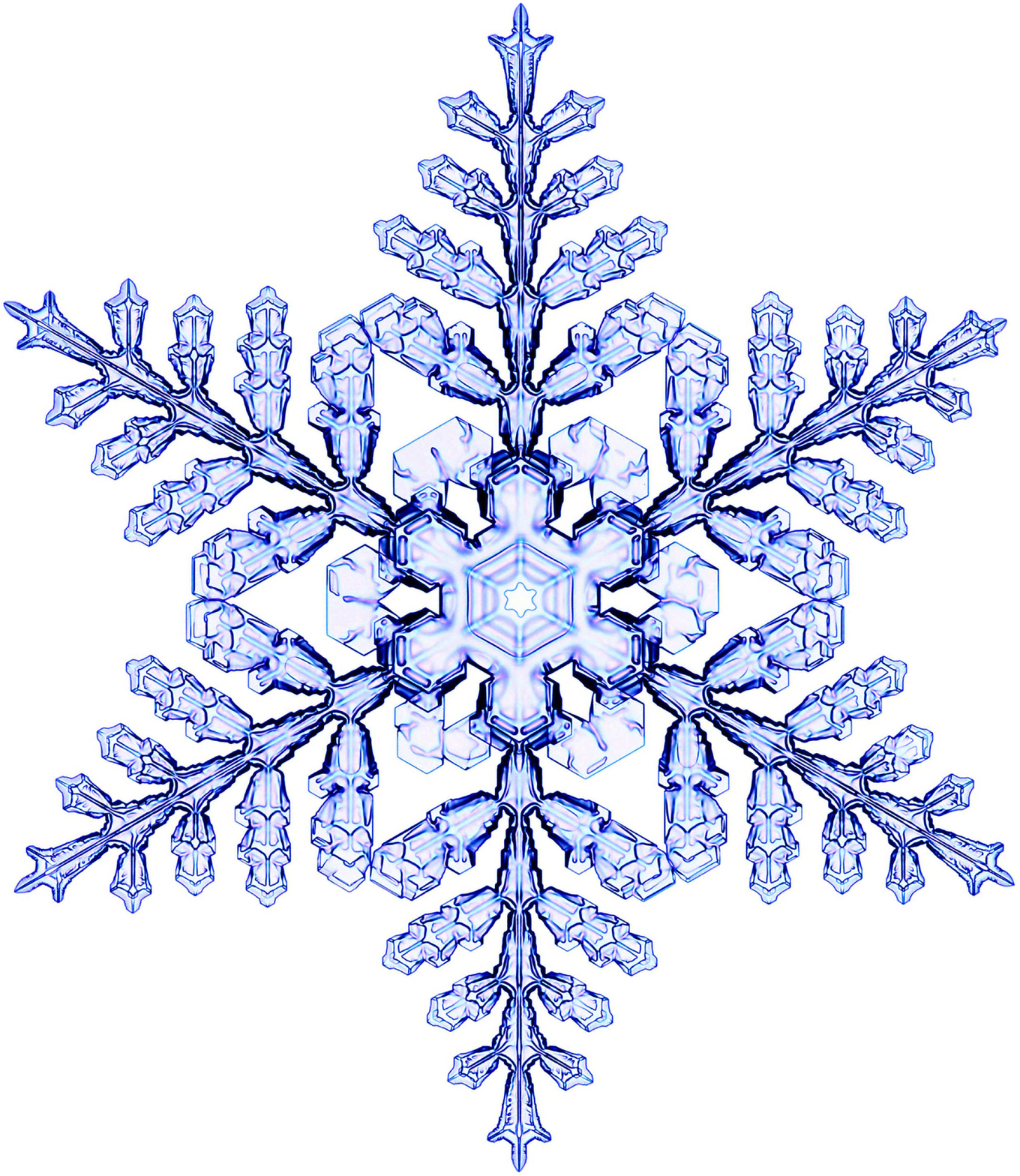}
  \caption{A dendritic POP crystal
measuring 3.6 mm from tip to tip. After forming the inner part of the
crystal, five sets of sidebranches were induced at regular intervals on the
six outer branches. Natural snow crystals rarely show this degree of
symmetry in their sidebranches.}
  \label{star}
\end{figure}

Figure \ref{sidebranching} shows an example of what we call an \textit{%
induced sidebranching event}. By first lowering the supersaturation to
produce faceting, and then increasing the supersaturation to stimulate
branching, it becomes possible to place sidebranches on a growing crystal in
a prescribed way. Figure \ref{star} shows another example were sidebranches
were placed five times at approximately uniform intervals along the main
branches. 

At still higher supersaturations, we might expect to see sidebranches
appearing spontaneously with irregular spacing along dendritic branches
growing near -15 C, as this is known to occur in diffusion-chamber studies
and in natural snow crystals. In the apparatus described here, however, we
have not found it possible to achieve sufficiently high supersaturations for
spontaneous sidebranching, because of the formation of water droplets on the
substrate.

\section{Discussion}

At present, our understanding of the diffusion-limited growth of structures
that are both branched and faceted remains quite poor. The cellular automata
method for numerical modeling of such structures holds much promise \cite%
{reiter05, gg09, kglca13, kelly13}, but we have not yet learned what input
physics should be applied in those models. We believe that
structure-dependent attachment kinetics (SDAK) is a key element to be
included \cite{sdak03, sdak12}, but many aspects of the SDAK hypothesis
remain untested. Detailed morphological comparisons between computational
models and ice growth experiments will likely play an important role in the
development of those models, especially in three dimensions. To this end,
ice crystals grown in the apparatus presented above should provide many
useful insights toward a better understanding of ice growth behavior in
particular, and of diffusion-limited crsytal growth more generally.

\bibliography{DSF1-Apparatus2-arXiv}
\bibliographystyle{unsrt}

\end{document}